# 4D ultrafast ultrasound imaging of naturally occurring shear waves in the human heart

C. Papadacci, V. Finel, O. Villemain, M. Tanter, M. Pernot

*Abstract*—The objectives were to develop a novel three-dimensional technology for imaging naturally occurring shear wave (SW) propagation, demonstrate feasibility on human volunteers and quantify SW velocity in different propagation directions. Imaging of natural SWs generated by valve closures has emerged to obtain a direct measurement of cardiac stiffness. Recently, natural SW velocity was assessed in two dimensions on parasternal long axis view under the assumption of a propagation direction along the septum. However, in this approach the source localization and the complex three-dimensional propagation wave path was neglected making the speed estimation unreliable. High volume rate transthoracic acquisitions of the human left ventricle (1100 volume/s) was performed with a 4D ultrafast echocardiographic scanner. Four-dimensional tissue velocity cineloops enabled visualization of aortic and mitral valve closure waves. Energy and time of flight mapping allowed propagation path visualization and source localization, respectively. Velocities were quantified along different directions. Aortic and mitral valve closure SW velocities were assessed for the three volunteers with low standard deviation. Anisotropic propagation was also found suggesting the necessity of using a three-dimensional imaging approach. Different velocities were estimated for the three directions for the aortic (3.4±0.1 m/s, 3.5±0.3 m/s, 5.4±0.7 m/s) and the mitral (2.8±0.5 m/s, 2.9±0.3 m/s, 4.6±0.7 m/s) valve SWs. 4D ultrafast ultrasound alleviates the limitations of 2D ultrafast ultrasound for cardiac SW imaging based on natural SW propagations and enables a comprehensive measurement of cardiac stiffness. This technique could provide stiffness mapping of the left ventricle.

## I. Introduction

MYOCARDIAL stiffness (MS) is a fundamental parameter for understanding and characterizing myocardial properties including contractility and relaxation, especially passive relaxation. Passive myocardial stiffness is an intrinsic biomechanical property of the myocardium that reflects the underlying myocardial structure and content at micro and macroscopic scales. Myocardial stiffness dysfunction is associated to severe cardiac pathologies. Heart failure with preserved ejection fraction (HFpEF) for instance, is associated to a high degree of risk of death [1] and its main pathophysiological effect is a passive myocardial stiffness increase [2]. Despite the urgent clinical need, there is no method available in the clinic to early detect HFpEF. Other cardiac pathologies would benefit from a direct measurement of cardiac stiffness. Myocardial stiffness mapping could be a major tool for characterization of post-infarction scars or to assess the extent of ablation zone in thermal ablation therapy.

In biomedical research, cardiac shear wave imaging has emerged to obtain a direct measurement of cardiac stiffness [3]. The feasibility of non-invasive myocardial stiffness assessment has been recently demonstrated on human patients [4]–[6]. It relies on the estimation of shear wave speed, a physical parameter directly linked to tissue stiffness [7]. To generate such shear waves, the source can be whether induced by an external force [3] or by an internal perturbation such as the valve closures [8], [9]. Cardiac shear wave imaging with externally induced acoustic radiation force presents some advantages and limitations. The main advantage is that MS can be measured at each point of the cardiac cycle, giving access to stiffness variation [10], contractility [3] or relaxation properties [11] in healthy humans [12], [13] and hypertrophic patients [4]. The main disadvantage is that it is only possible to analyze a specific myocardial region, as defined by the push area. The full chamber stiffness assessment in one cardiac cycle is not possible using this approach. However, shear wave imaging

Manuscript received X
This study was supported by the European Research Council under the European Union's Seventh Framework Program (FP/2007-2013) / ERC Grant Agreement n° 311025, the ANR-10-IDEX-0001-02 PSL* Research University and the Leducq Foundation in the framework of RETP program. We acknowledge the ART (Technological Research Accelerator) biomedical ultrasound program of INSERM. The Titan X Pascal used for this research was donated by the NVIDIA Corporation.

C.Papadacci, V.Finel, O.Villemain, M.Tanter, M.Pernot are with Physics for Medicine laboratory, ESPCI Paris, Inserm, CNRS, PSL (e-mail: clement.papadacci@espci.fr).

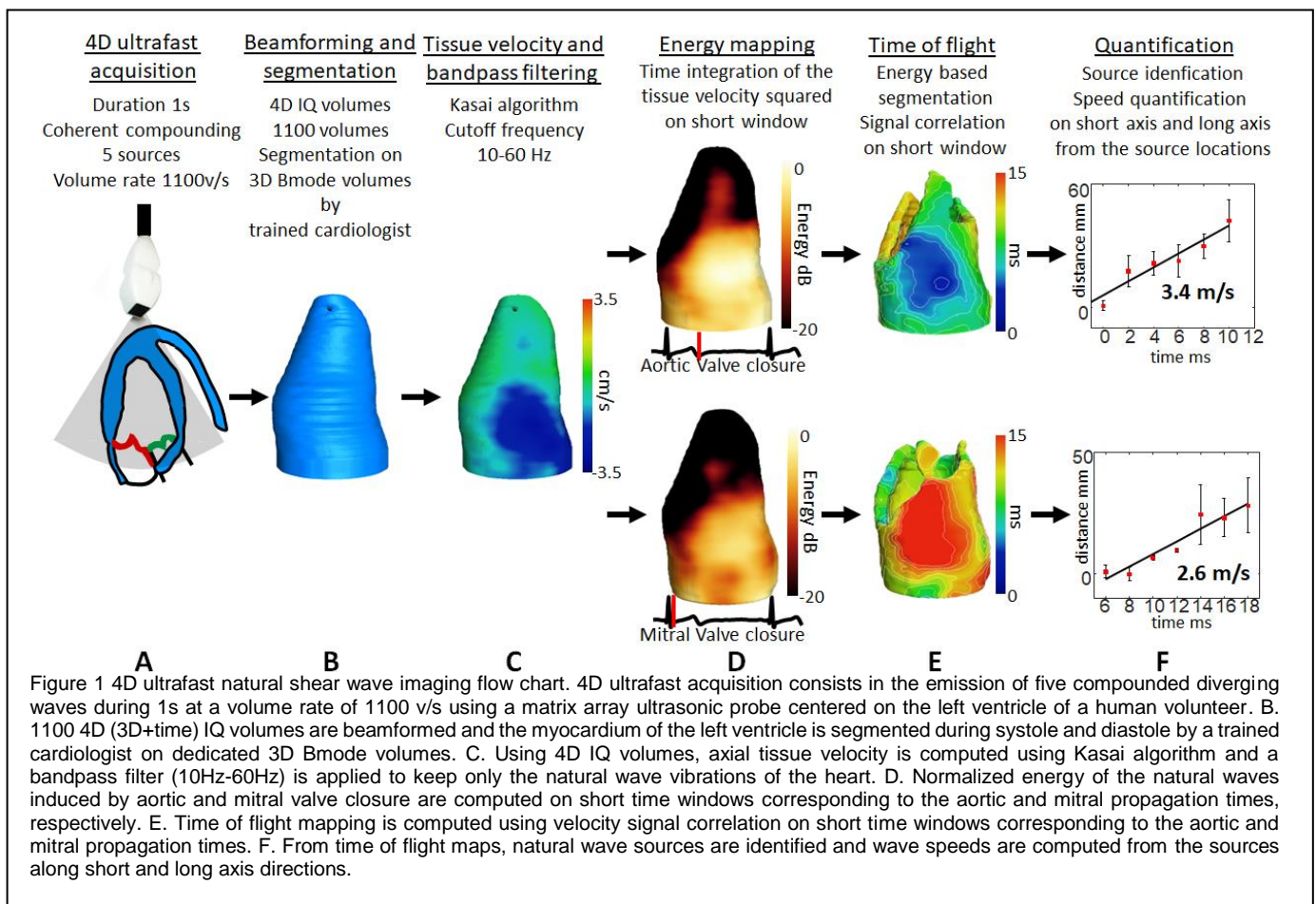

Figure 1 4D ultrafast natural shear wave imaging flow chart. 4D ultrafast acquisition consists in the emission of five compounded diverging waves during 1s at a volume rate of 1100 v/s using a matrix array ultrasonic probe centered on the left ventricle of a human volunteer. B. 1100 4D (3D+time) IQ volumes are beamformed and the myocardium of the left ventricle is segmented during systole and diastole by a trained cardiologist on dedicated 3D Bmode volumes. C. Using 4D IQ volumes, axial tissue velocity is computed using Kasai algorithm and a bandpass filter (10Hz-60Hz) is applied to keep only the natural wave vibrations of the heart. D. Normalized energy of the natural waves induced by aortic and mitral valve closure are computed on short time windows corresponding to the aortic and mitral propagation times, respectively. E. Time of flight mapping is computed using velocity signal correlation on short time windows corresponding to the aortic and mitral propagation times. F. From time of flight maps, natural wave sources are identified and wave speeds are computed from the sources along short and long axis directions.

relying on natural shear waves generated by valve closures could enable a more global MS assessment. The natural valve waves are generated by strong low frequency mechanical events and propagate in the heart over large distances. However, natural shear wave propagation in the myocardium is a complex tri-dimensional phenomenon. By nature, its source is uncontrolled in terms of location, wavelength and amplitude. The propagation occurs in three dimensions along unknown directions. As a consequence, quantification of the stiffness from shear wave propagation requires a complex three-dimensional inverse problem. A first aortic valve closure wave speed measurement was performed by Kanai in 2005 [14]. In most of the recent studies, the natural wave velocities were assessed in two dimensions on the parasternal long axis view under the assumption of a propagation direction along the septum [15][5]. This approach is based on a strong assumption that have not yet been verified: the shear wave propagation is assumed to be only two-dimensional. However, the precise source location remains unidentified and the estimated velocity along the septum could simply be a projection of the absolute speed as the propagation occurs in three dimension. Physiological variability among patients in terms of valve geometry and heart morphology could thus induce different propagation directions making the velocity estimation not reliable. In order to present a reliable assessment of the MS, there is therefore a strong need for a 3D approach to precisely localize natural wave sources to track them along different directions.

Recently, 4D ultrafast ultrasound imaging has emerged. It enables the acquisition of entire volumes at very high volume rates (>5,000volumes/s) [16]. In cardiac applications, it has already been applied to cardiac fiber tractography [17], cardiac strain for lesion detection [18] and recently to the semi-automatic assessment of cardiac Doppler indices [19]. It has also been used to track shear waves [20] and fast tissue motion [19] in entire volumes.

In this study, we propose to apply 4D ultrafast imaging to the tracking of natural shear waves in the human heart, to estimate myocardial stiffness in 3D. The objectives of the study were to 1) develop a methodology for 3D shear wave propagation imaging 2) demonstrate the feasibility on human volunteers and 3) quantify the shear wave velocities in different propagation directions.

## II. METHODS

The following method description has been illustrated in the flow chart of Figure 1 and Figure 2.
1. Experimental set-up
    a. Experiment

Experiments were performed on three healthy volunteers, male, 27, 28 and 30 years old. The volunteers were placed in left decubitus position and scanned by a trained cardiologist. 3D probe was positioned on apical view centered on left ventricle using two dimensional real-time B-mode images. Ultrafast acquisition was triggered on the R-wave from electrocardiogram (ECG). 3D Bmode sequence triggered on the R-wave followed the UF acquisition.

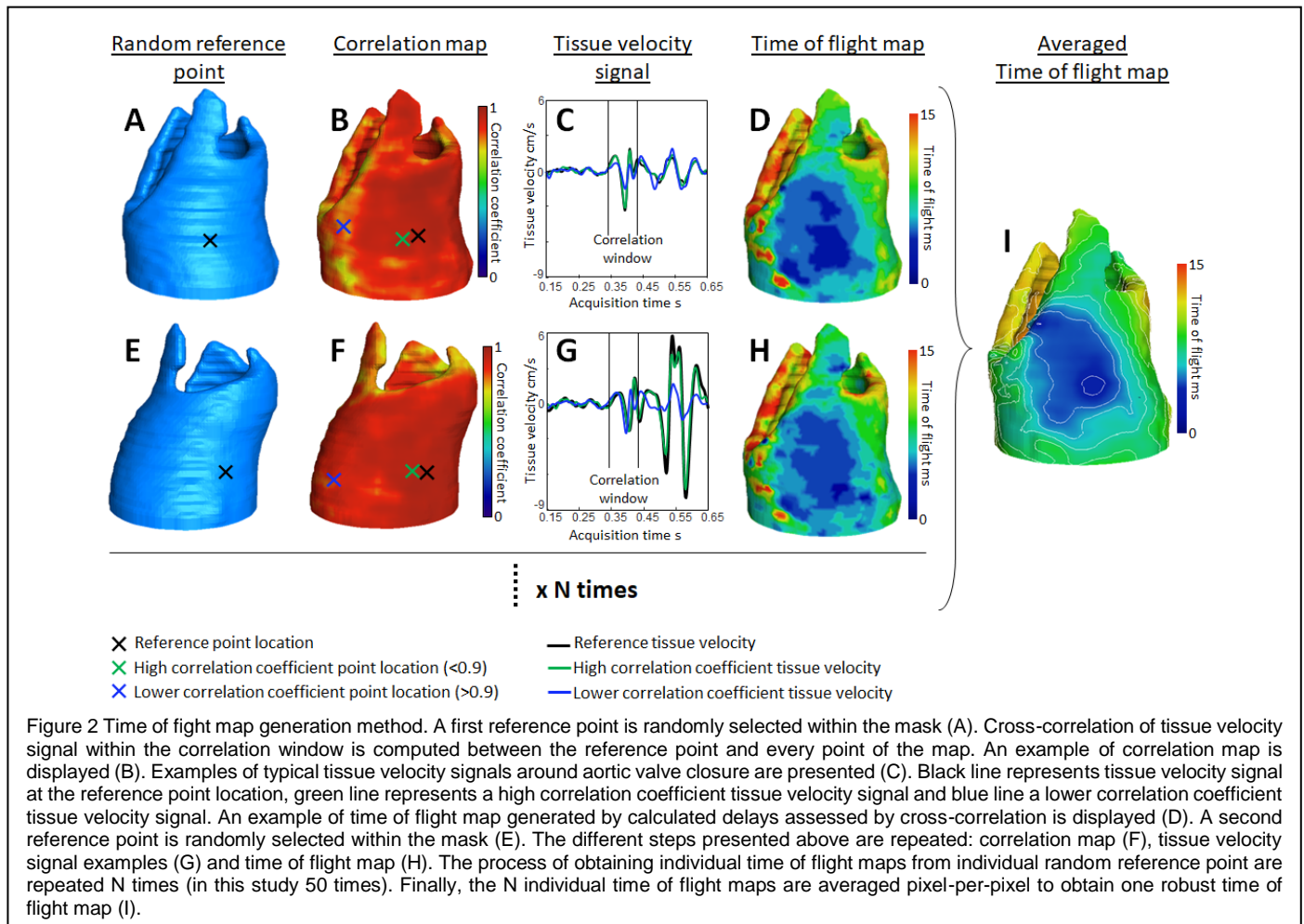

Figure 2 Time of fight map generation method. A first reference point is randomly selected within the mask (A). Cross-correlation of tissue velocity signal within the correlation window is computed between the reference point and every point of the map. An example of correlation map is displayed (B). Examples of typical tissue velocity signals around aortic valve closure are presented (C). Black line represents tissue velocity signal at the reference point location, green line represents a high correlation coefficient tissue velocity signal and blue line a lower correlation coefficient tissue velocity signal. An example of time of flight map generated by calculated delays assessed by cross-correlation is displayed (D). A second reference point is randomly selected within the mask (E). The different steps presented above are repeated: correlation map (F), tissue velocity signal examples (G) and time of flight map (H). The process of obtaining individual time of flight maps from individual random reference point are repeated N times (in this study 50 times). Finally, the N individual time of flight maps are averaged pixel-per-pixel to obtain one robust time of flight map (I).

ECG was co-recorded during the acquisition.

    b. Materials

Acquisitions were performed using a 3MHz matrix array probe of 1024 elements (Vermon, France) connected to a 4D ultrafast ultrasound device (Verasonics, US) with 1024 electronic channels in emission and receive. An electrocardiogram device (Accusync, US) with R-wave output was connected to the trigger clock of the 4D ultrafast ultrasound device.

  2. Data acquisition
    a. Ultrasonic emission
      i. Ultrafast acquisition

5500 diverging waves were emitted from the 3D probe during 1s at a pulse repetition frequency of 5500 Hz in which 5 different sources were alternatively emitted to allow coherent compounding as it is performed in [16], [21]. Effective frame rate was 1100 volumes/s after coherent compounding. Emission central frequency was set to 3 MHz. In this study, no harmonic imagine was used.

      ii. Bmode acquisition

5500 diverging waves were emitted from the 3D probe during 1s at a pulse repetition frequency of 5500 Hz in which 25 different sources were alternatively emitted to allow coherent compounding. Effective frame rate was 220 volumes/s after coherent compounding.

    b. Ultrasonic receive

For each diverging wave emission, backscatter echoes were digitalized at a sampling frequency of 12 MHz and recorded in memory.

  3. Data processing
    a. Beamforming and segmentation

3D delay-and-sum beamforming including coherent compounding were performed to reconstruct 1100 In-phase and Quadrature (IQ) volumes for the UF sequence and 200 IQ volumes for the Bmode sequence. Volume lateral dimensions were set to 80° by 80° by 12 cm depth. Pixel dimensions were set to 1° by 1° by 610 μm. 3D B-mode volumes were computed by calculating the absolute value of the IQ volumes and by using log compression. Example of two orthogonal slices from a 3D Bmode volume is displayed in Figure 3 A,B.

At end-systolic and end-diastolic times, 3D Bmode volumes were extracted. Segmentation was performed by a trained cardiologist on 2D slices generated every 5mm to compute 3D aortic and mitral valve closure masks, respectively. An example of segmented slices at end-systolic time was displayed in Figure 3 C,D,E.

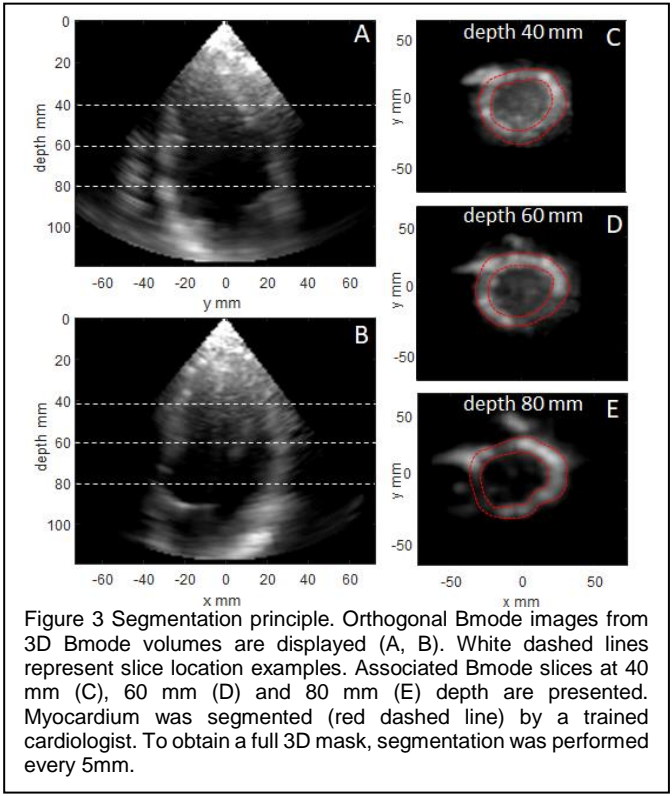

Figure 3 Segmentation principle. Orthogonal Bmode images from 3D Bmode volumes are displayed (A, B). White dashed lines represent slice location examples. Associated Bmode slices at 40 mm (C), 60 mm (D) and 80 mm (E) depth are presented. Myocardium was segmented (red dashed line) by a trained cardiologist. To obtain a full 3D mask, segmentation was performed every 5mm.

b. Tissue velocity and bandpass filtering

Axial tissue velocities were computed from axial cross-correlation of volume to volume IQ data (Kasai algorithm). A Butterworth bandpass filter with a cutoff frequency of 10 Hz to 60 Hz was applied to the tissue velocity.

c. Time window identification

Aortic and mitral valve closure timings were manually identified by a trained cardiologist using the ECG. A time window of 90ms was defined around these events which was used as a correlation kernel.

d. Energy mapping

Squared 4D axial tissue velocity was computed and integrated over the two time windows. The resulting energy was then normalized by the maximum energy in the map. A logarithmic compression was applied. Normalized energy maps were displayed in decibels. The energy map $\varepsilon(x,y,z)$ was defined as:

$$\varepsilon(x,y,z) = 20 log10 \left( \frac{\int_{t_0}^{t_1} v^2(x,y,z,t)}{\max_{x,y,z} \left( \int_{t_0}^{t_1} v^2(x,y,z,t) \right)} \right)$$

Where: $v(x,y,z,t)$ is the 4D axial tissue velocity, $t_1 - t_0$ is the correlation window and $\max_{x,y,z}$ is maximum detection over 3D space.

A -20dB threshold was set arbitrarily from the energy maps to remove from the original mask the regions where the wave did not propagate.

e. Time of flight

Figure 2 illustrates the post-processing steps to compute time of flight map.

A reference point at a random location was first selected in the final mask region (Figure 2 A). Cross-correlation within short time window was performed between tissue velocity of reference point and every points in the mask (Figure 2 C) enabling time delay assessment. Correlation coefficient values were calculated and mapped as an example in Figure 2 B. Time delays were mapped in three dimensions to obtain a single time of flight map (Figure 2 D). Another reference point at a random location was then selected in the final mask region (Figure 2 E), cross-correlation was performed between the reference point and every points in the mask (Figure 2 G) to assess time delays and correlation coefficient values (Figure 2 F). A second time of flight map was then computed (Figure 2 H). The process described above was repeated 50 times to finally obtain 50 cross-correlation value maps and 50 time of flight maps. For each time of flight map, the source was automatically identified as the minimum time delay into the map and the value t=0ms was set to this location. The 50 rescaled time of flight maps were averaged pixel-to-pixel to obtain a single time of flight map (Figure 2 I). Isosurfaces every 2ms were computed and displayed to improve visualization. A single cross-correlation coefficient value was obtained by averaging the 50 cross-correlation maps pixel-to-pixel and averaging this final cross-correlation map over space.

f. Quantification

The full 3D inverse problem of the shear wave propagation in the left ventricle remains a complex development and is out of the scope of this study. Therefore, a simple estimation approach was used to determine the wave velocity depending on the propagation direction. The approach consisted in determining the wave source location on the time of flight map (to=0ms) and in using a linear regression of the time of flight along 1D curves set in the long axis and short axis directions. 1D curves were displayed as the averaged distance as a function of discrete time 0ms, 2ms, 4ms etc and standard deviation over distance was displayed.

g. Shear wavelength

Natural shear wave central frequencies (f) were estimated as the maxima of the averaged Fourier transforms of tissue velocity signals within the valve time windows from 50 points randomly located in the myocardium. Shear wavelengths were computed along three directions as v/f where v is the natural wave velocities.

h. Processing materials

Beamforming was performed on GPU units (Titan XP, NVidia, US). Processing was performed with Matlab software. 3D representations were computed with Amira software.

III. RESULTS

1. Aortic valve closure wave (end-systole)

a. Propagation from tissue velocity

The propagation of the shear wave produced by the aortic valve closure was observed in Figure 4 on the filtered 4D axial tissue velocity at the end of the ventricular repolarization, Figure 4.C. The wave source was identified near the base of the heart (Figure 4.A), on the antero-septal segment (Figure 4.B). This particular location corresponds to the physiologic aortic valve position in the heart. The wave propagated along the anterior wall and the septum to finally reach the posterior wall (Figure 4.A). From the cross-section view (Figure 4.B), the wave propagated faster from the septum to the posterior wall, than from the septum to the anterior wall. Total propagation duration lasted less than 20 ms. For a better visualization cineloops are attached to (Figure 4).

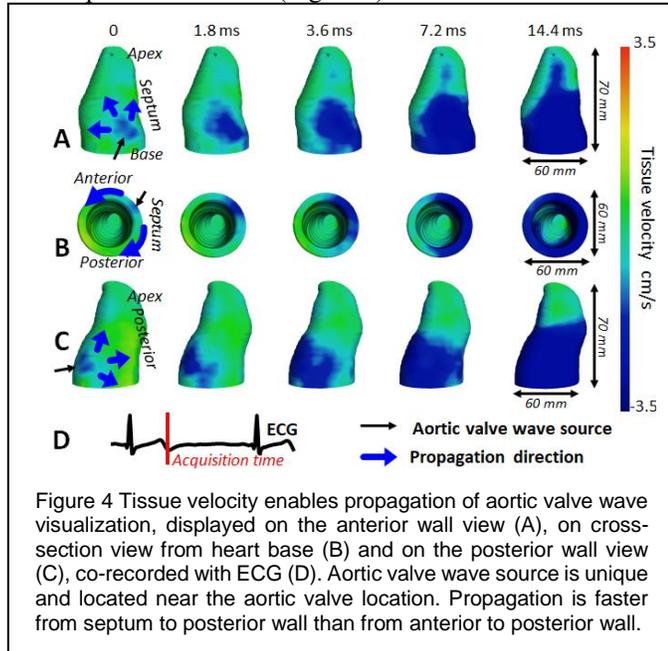

Figure 4 Tissue velocity enables propagation of aortic valve wave visualization, displayed on the anterior wall view (A), on cross-section view from heart base (B) and on the posterior wall view (C), co-recorded with ECG (D). Aortic valve wave source is unique and located near the aortic valve location. Propagation is faster from septum to posterior wall than from anterior to posterior wall.

### b. Energy and time of flight mapping

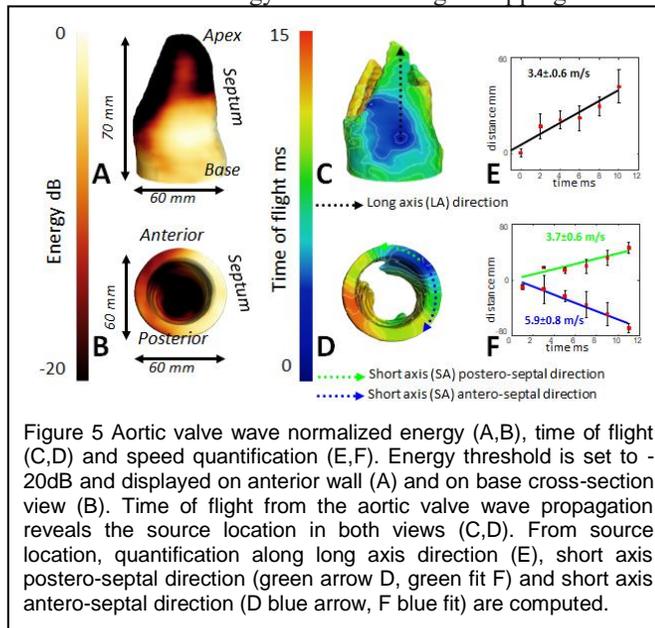

Figure 5 Aortic valve wave normalized energy (A,B), time of flight (C,D) and speed quantification (E,F). Energy threshold is set to -20dB and displayed on anterior wall (A) and on base cross-section view (B). Time of flight from the aortic valve wave propagation reveals the source location in both views (C,D). From source location, quantification along long axis direction (E), short axis postero-septal direction (green arrow D, green fit F) and short axis antero-septal direction (D blue arrow, F blue fit) are computed.

Normalized energy of the aortic valve wave propagation was computed (Figure 5.A,B). A threshold was set to -20dB, meaning that no wave propagation was considered in heart regions under this particular value (equation was provided in the method section). Larger and smaller energy were measured at the wave source and the apex respectively, as displayed in Figure 5.A,B. Time of flight maps were then computed. Blue and red colors represent early and late propagation times, respectively. As observed on the wave propagation frames (or cineloop), a unique source was identified near the base of the heart (Figure 5.C), on the antero-septal segment (Figure 5.D). From the cross-section view (Figure 5.D), the wave propagated faster from the septum to the posterior wall, than from the septum to the anterior wall. Indeed, the red color indicates that for the same arrival time, the wave covers more distance when it propagates on the right direction (blue arrow) than on the left direction (green arrow).

Wave velocity quantification was performed from the source along three directions: long axis (black arrow Figure 5.C,E), short axis postero-septal (green arrow Figure 5.D,F) and short axis antero-septal (blue arrow Figure 5.D,F). We found 3.4±0.6 m/s, 3.7±0.6 m/s and 5.9±0.8 m/s for long axis, short axis left and short axis right directions, respectively.

### c. Time of flight mapping and cross-correlation coefficient values for the three volunteers

The experiment was performed on three volunteers. Similar results were found as shown in Figure 6 and table 1.

For each volunteer, high cross-correlation coefficient values were found (table 1 average of 0.83±0.02).

A unique source was identified near the base of the heart (Figure 6.A,B,C), on the antero-septal segment (Figure 6.D,E,F). Similar patterns were observed with a faster propagation along the right direction than the left. Wave velocities were quantified and summarized in Table 1.

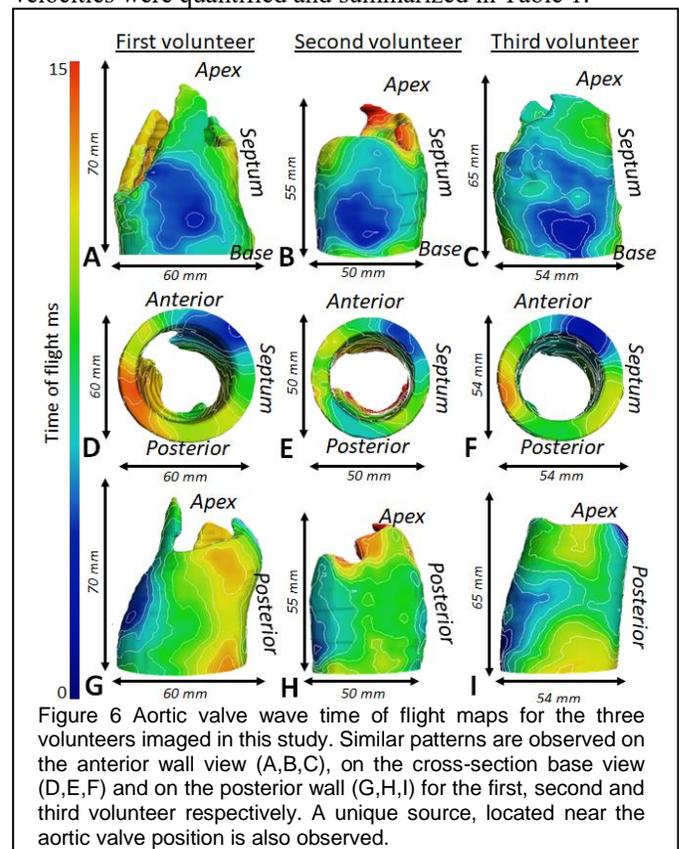

Figure 6 Aortic valve wave time of flight maps for the three volunteers imaged in this study. Similar patterns are observed on the anterior wall view (A,B,C), on the cross-section base view (D,E,F) and on the posterior wall (G,H,I) for the first, second and third volunteer respectively. A unique source, located near the aortic valve position is also observed.

Natural wave velocities were quantified for each volunteer along the long axis and short axis directions from the identified sources. Speed was averaged over the three volunteers and standard deviation was calculated over these results. For aortic valve closure wave, speeds of 3.4±0.1 m/s, 3.5±0.3 m/s and 5.4±0.7 m/s were found for long axis, short axis postero-septal and antero-septal directions, respectively. Average wavelength of 0.17±0.01m, 0.18±0.02m and 0.27±0.04m were found for long axis, short axis postero-septal and antero-septal directions, respectively.

| | Aortic valve | | | | | | Left Ventricle |
|---|---|---|---|---|---|---|---|
| | LA | | SA antero-septal | | SA postero-septal | | |
| | Wave speed (m/s) | Wavelength (m) | Wave speed (m/s) | Wavelength (m) | Wave speed (m/s) | Wavelength (m) | Correlation coefficient |
| First volunteer | 3.4±0.6 | 0.17 | 3.7±0.6 | 0.19 | 5.9±0.8 | 0.3 | 0.81±0.15 |
| Second volunteer | 3.3±0.6 | 0.17 | 3.2±0.5 | 0.16 | 5.7±0.7 | 0.29 | 0.85±0.15 |
| Third volunteer | 3.4±0.5 | 0.16 | 3.7±0.5 | 0.18 | 4.6±0.6 | 0.22 | 0.84±0.16 |
| Average | 3.4 | 0.17 | 3.5 | 0.18 | 5.4 | 0.27 | 0.83 |
| Standard dev | 0.1 | 0.01 | 0.3 | 0.02 | 0.7 | 0.04 | 0.02 |

Table 1. Natural wave speed and wavelength results for the three volunteers. The aortic valve wave speeds and wavelengths are calculated along long axis (LA) direction, short axis (SA) postero-septal, and antero-septal directions. Mean Correlation coefficient from cross-correlation algorithm is also displayed.

2. Mitral valve closure wave (end-diastole)

a. Propagation from tissue velocity

The shear wave corresponding to mitral valve closure was observed at the onset of the ventricular contraction on the filtered 4D axial tissue velocities (Figure 7.D). The spatial extent of the source was found to be larger compared to the aortic valve closure source. Two wave sources, S1 and S2, were identified near the base of the heart (Figure 7.A,C). S1 is located on the antero-septal segment (Figure 7.A,B) and S2, on the posterior wall (Figure 7.B,C). S2 appeared to be 5 to 10 ms earlier than S1. These particular locations may correspond to the motion generated by the two physiologic mitral valve leaflets. The waves propagated from these sources along the anterior (Figure 7.A) and posterior wall (Figure 7.C). On the cross-section view (Figure 7.B), the principal propagation direction is from the anterior wall to the posterior wall through the septum. Total propagation duration lasted less than 20 ms. For a better visualization, cineloops are attached to (Figure 7).

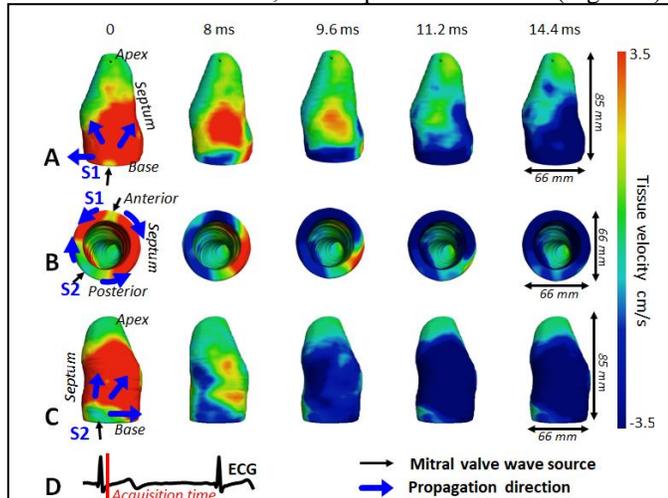

Figure 7 Tissue velocity enables mitral valve wave propagation visualization, displayed on the anterior wall view (A), on cross-section view from heart base (B) and on posterior wall view (C), co-recorded with ECG (D). Mitral valve wave sources are dual (S1 and S2) and located around the mitral valve's leaflets. Clear propagation patterns are observed along the long axis of the left ventricle. On cross-section view from heart base, propagation from S1 to posterior wall through the septum is clearly observed.

b. Energy and time of flight mapping

Normalized energy of the mitral valve wave propagation was computed (Figure 8.A,B,C). A threshold was set to -20dB, meaning that we did not consider the wave propagation in heart regions under this particular value. Similarly to the aortic valve wave, larger and smaller energy were measured at the wave source and the apex, respectively, as displayed in Figure 8.A and C.

Time of flight maps were then computed. Blue and red colors represent early and late propagation time, respectively. As observed on the wave propagation frames (or cineloop), two sources were identified near the base of the heart (Figure 8.D,F), the first source S1, on the antero-septal segment and a second early source S2 on the posterior wall (Figure 8.E). From the cross-section view (Figure 8.E), the wave propagated mainly from source S1 to source S2 (blue arrow), which was confirmed from the cineloop of the propagation.

Wave speed quantification was performed from the source along three directions: S1 long axis (black arrow Figure 8.D,G), S1 short axis (blue arrow Figure 8.E,H) and S2 long axis (green arrow Figure 8.F,I). We found 2.6±0.6 m/s, 4.0±0.8 m/s and 2.8±0.7 m/s for S1 long axis, S1 short axis and S2 long axis directions, respectively.

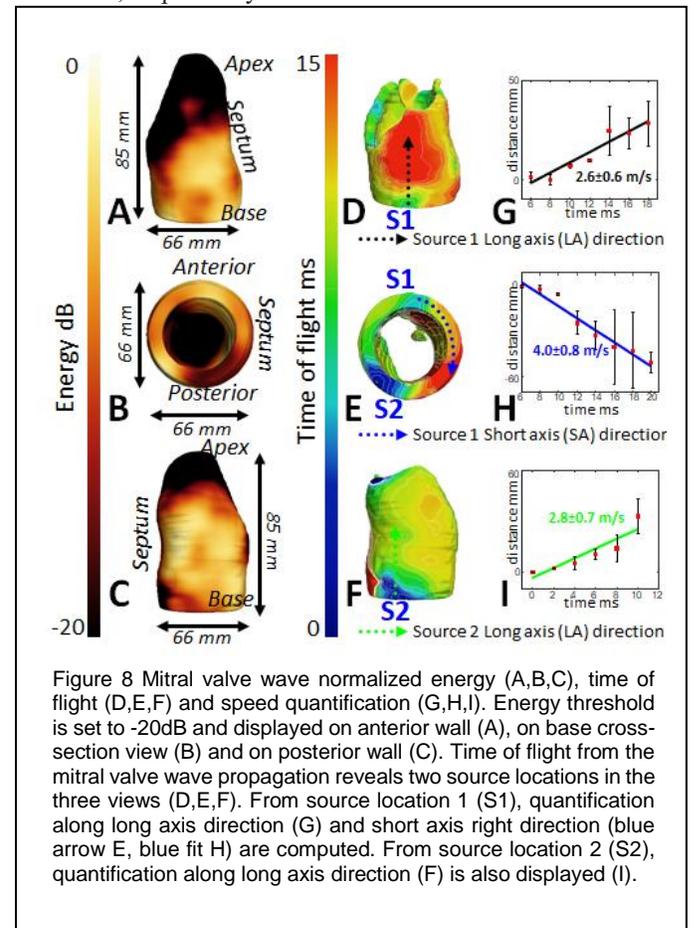

Figure 8 Mitral valve wave normalized energy (A,B,C), time of flight (D,E,F) and speed quantification (G,H,I). Energy threshold is set to -20dB and displayed on anterior wall (A), on base cross-section view (B) and on posterior wall (C). Time of flight from the mitral valve wave propagation reveals two source locations in the three views (D,E,F). From source location 1 (S1), quantification along long axis direction (G) and short axis right direction (blue arrow E, blue fit H) are computed. From source location 2 (S2), quantification along long axis direction (F) is also displayed (I).

c. Time of flight mapping and cross-correlation coefficient values for the three volunteers

The same experiment was performed on three volunteers. Similar results were found as shown in Figure 9.

## IV. DISCUSSION

In this study, we demonstrated the feasibility of estimating myocardial shear velocities in 3D along different directions using 4D natural shear wave imaging of the human heart. Furthermore, the complex propagation of both waves induced by aortic and mitral valve closure was observed for the first time in four dimensions (3D volume+time) by assessing local axial tissue velocity changes. Energy maps representing regions where the waves propagated were computed to increase our understanding of propagation path and to obtain a mask. Time of flight maps and isolines enabled graphical representation of propagation and source location detection. From source locations, shear wave velocities were assessed along different directions enabling quantifications.

In this study, we introduced several novel methods to quantify the shear wave velocity in the entire left ventricle: The first step of our post-processing approach relied on the quantification of the time of flight of shear wave propagation using cross-correlation of short time window between tissue velocity signal of reference point and every points in the myocardium. The average over 50 individual time of flight maps generated from 50 random reference points insured robustness. With this method, a complete and robust time of flight map was achieved at every voxel of the myocardium. In a second step, an automatic source identification algorithm was used to rescale the time maps before averaging them. Relevance of the method was demonstrated by high averaged cross-correlation coefficient values found for each volunteer. Furthermore, cross-correlation coefficient map examples showed high cross-correlation coefficient values everywhere in the mask for individual random reference points as no distance dependency was found. The last innovative step of the post-processing consisted in analyzing the shear velocity in different directions from the source point location.

Natural wave speeds and wavelengths were successfully assessed in the three human volunteers. For aortic valve closure, a unique source was observed and localized in the septum near the aortic valve location. The shear velocity was assessed from base to apex direction (LA direction). In this case, the LA direction was consistent with the direction assessed in two dimensional ultrafast imaging using parasternal long axis view in the recent studies [5], [22]. Speed values were found in good agreement with literature [22]. Cross section view (SA) of the myocardium allowed quantifications from the source to the left and the right directions. Interestingly, the wave propagated faster along the antero-septal direction than the postero-septal direction for each volunteer. Wave guidance effects along a different wall structures could be an explanation of this phenomenon. Moreover, shear wave propagated faster along SA than LA. This could be due to the fact that shear waves propagate faster along cardiac fibers that are mainly distributed along SA in the wall. This anisotropy which has been observed with natural waves for the first time in this study, has already been described in literature regarding remotely induced shear wave propagation [23].

For the wave generated at the mitral valve closure, two sources were observed and localized. One source was located in the antero-septal wall. This observation was consistent with two dimensional ultrafast acquisition in parasternal long axis

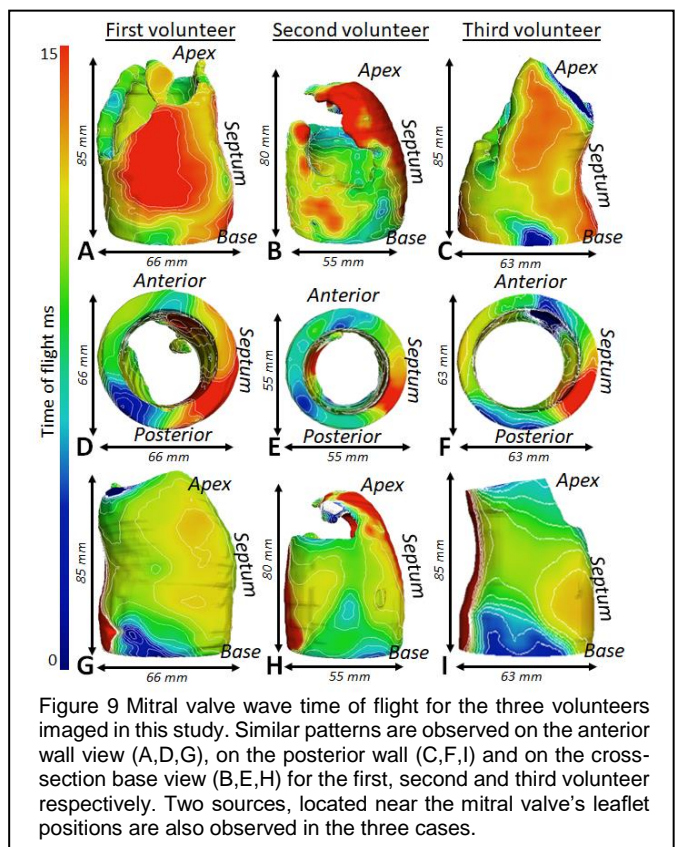

Figure 9 Mitral valve wave time of flight for the three volunteers imaged in this study. Similar patterns are observed on the anterior wall view (A,D,G), on the posterior wall (C,F,I) and on the cross-section base view (B,E,H) for the first, second and third volunteer respectively. Two sources, located near the mitral valve's leaflet positions are also observed in the three cases.

For each volunteer, high cross-correlation coefficient values were found (table 2 average of 0.78±0.04). Two sources were identified near the base of the heart as displayed in Figure 9.A,B,C for source 1, and in Figure 9.G,H,I for source 2. On the antero-septal segment for S1 and on the posterior wall for S2. Wave velocities were quantified and summarized in Table 2.

The wave velocities were quantified for each volunteer along different directions from the identified sources. Velocities were averaged over the three volunteers and standard deviations were calculated over these results. For mitral valve closure wave, velocities of 2.8±0.5 m/s, 2.9±0.3 m/s and 4.6±0.7 m/s were found for long axis source 1, long axis source 2 and short axis source 1 directions, respectively. Average wavelengths of 0.14±0.02 m, 0.16±0.02 m and 0.25±0.04 were found for mitral long axis source 1, long axis source 2 and short axis source 1 directions, respectively.

|  | Mitral valve | | | | | | Left ventricle |
|---|---|---|---|---|---|---|---|
|  | LA source 1 | | LA source 2 | | SA source 1 | | |
|  | Wave speed (m/s) | Wavelength (m) | Wave speed (m/s) | Wavelength (m) | Wave speed (m/s) | Wavelength (m) | Correlation coefficient |
| First volunteer | 2.6±0.6 | 0.14 | 2.8±0.7 | 0.16 | 4±0.8 | 0.22 | 0.73±0.19 |
| Second volunteer | 3.4±0.7 | 0.19 | 3.2±0.5 | 0.18 | 5.3±0.7 | 0.29 | 0.81±0.19 |
| Third volunteer | 2.5±0.6 | 0.13 | 2.6±0.7 | 0.14 | 4.4±0.9 | 0.24 | 0.80±0.20 |
| Average | 2.8 | 0.14 | 2.9 | 0.16 | 4.6 | 0.25 | 0.78 |
| Standard dev | 0.5 | 0.02 | 0.3 | 0.02 | 0.7 | 0.04 | 0.04 |

Table 2. Natural wave speed and wavelength results for the three volunteers. The mitral valve wave speeds and wavelengths are calculated along long axis (LA) direction for source 2, long axis and short axis (SA) right for source 1. Mean Correlation coefficient from cross-correlation algorithm is also displayed.

view used to assess the mitral valve shear wave speed [22]. The other identified source was located on the other side of the heart, on the left posterior cardiac wall. The existence of these two sources could be explained by the size, the geometry and the location of the mitral valve. Indeed, the mitral valve diameter is significantly larger than the diameter of the aortic valve, it is composed of two leaflets (or flaps) that are 'hooked' to the left posterior wall and to the antero-septal segment. The interpretation is that when the valve closes, two waves are generated by the two leaflets that first appear in the myocardium where the valve attaches. Shear velocity quantification was performed along three directions. In LA direction, speed was quantified from the first and second source exhibiting similar results. LA speed from the first source was in good agreement with the results found in literature [5], [22]. On the cross-section view, clear propagation was only observed from the first source to the septum. On the other side of the heart the rapid mixture between waves coming from source 1 and 2 did not allow quantification. These complex interferences can be observed on cineloop attached to the manuscript. Interferences on such small distances did not allow accurate shear wave speed calculations as the propagation path remains challenging to determine for each wave. On another hand, propagation from source 2 to source 1 over the septal wall was not observed. It could be due to physiological blockade on this specific region of the ventricle. SA speed was found to be higher than LA speed probably due to fiber direction.

Overall, aortic valve closure wave velocities were found to be lower in average than the mitral valve ones. This observation was found in most of the studies in literature [5].

Schematic representation of aortic (A) and mitral valve (B) closure wave displayed from heart base are shown in Figure 10.

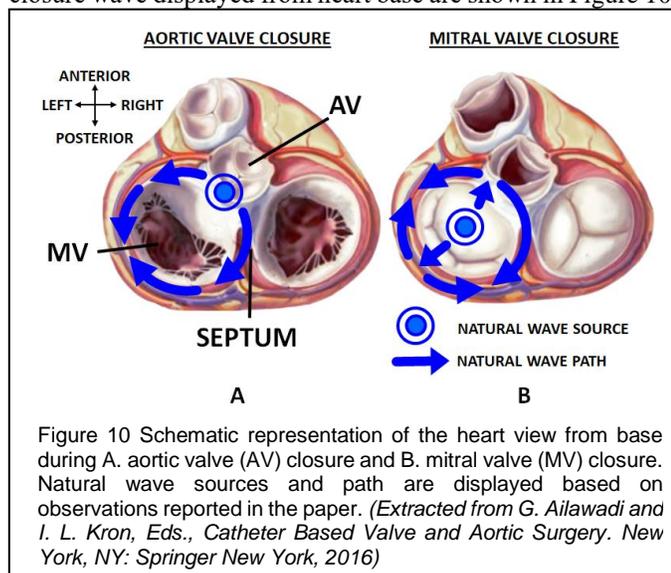

Figure 10 Schematic representation of the heart view from base during A. aortic valve (AV) closure and B. mitral valve (MV) closure. Natural wave sources and path are displayed based on observations reported in the paper. *(Extracted from G. Ailawadi and I. L. Kron, Eds., Catheter Based Valve and Aortic Surgery. New York, NY: Springer New York, 2016)*

This new method alleviated most of the limitations inherent to 2D investigations. The natural wave source and complex propagation paths were identified allowing measurement of vector valued velocities of the waves emanating from the sources. The method removed the risk of assessing a projection of the true speed as it is done in 2D method as the source is not necessarily contained in the field of view. This study is also significant as it increases our knowledge of cardiac physiology at the millisecond scale. 4D imaging could also increase robustness and reproducibility of shear wave tracking by assessing more than one speed at the source location. Finally, it is the first step towards new imaging biomarkers of the ventricular stiffness by mapping the shear velocity over a large part of the ventricle. Quantitative 3D mapping of the shear velocity will require, however, resolving a complex inverse problem of 3D shear wave propagation in the myocardium. Physiological and physical natural wave parameters such as ventricle geometry, wall thickness, myofiber anisotropy and viscosity would require complex modelling to solve the inverse problem. The development of such a model goes beyond the scope of this manuscript.

With 4D ultrafast imaging a velocity profile curve was retrieved from each voxel. Figure 2C,G showed an example at three points of space. For the three volunteers, similar patterns were observed. Central frequency (around 20 Hz) was found to be in the same order of magnitude compared to the results found in the literature (around 25Hz in [9], around 30-47Hz in [22]). The differences could be explained by the difference in the chosen time window size in agreement to the apparent time duration (around 60ms in [22], 75ms in [9]).

There are still interrogations inherent to the natural shear waves. For instance, the tricuspid valve closes quasi-simultaneously with the mitral valve and is also directly attached to the septum. The identification of the closure exact timing and the localization of the tricuspid valve closure source(s) are currently an ongoing work. From tricuspid valve wave propagation, shear wave speed could be assessed in the right ventricle allowing a complete mapping of the heart during isovolumic contraction.

Moreover, aortic and mitral valve closures occur right before isovolumic relaxation and contraction respectively, two rapid events occurring within less than 40 ms which is the same order of magnitude than the natural wave propagation duration. It is thus challenging to know exactly at which state is the heart at these very particular moments as the stiffness varies rapidly [24].

Another interrogation remains in the physiological meaning of low energy regions. For all the volunteers, in the apex and the upper free wall, the valve wave energy was below -20dB. Displacement in the apex is complex and the axial velocity component could be small compared to other velocity components. Small energy could also be due to reverberation clutter from fat and muscular tissues above the apex.

In this study, shear wave speed assessment was based on axial displacement estimation only. Other components of the shear wave speed may be interesting to study to better understand valve wave nature and could provide a better description of the shear wave propagation.

The goal of this study was to demonstrate the feasibility of visualizing naturally occurring shear waves in the heart in 3D and estimating their speed. Young volunteers were chosen to ensure healthy hearts and consistency among volunteers. Reproducibility study on healthy volunteers and a clinical study on heart failure patients are ongoing works.

50 individual time of flight maps from 50 reference points randomly selected to compute a single time of flight map were used to reduce variability and increase robustness. However, this number has not been optimized and a smaller number may be sufficient to achieve similar results.

3D ultrafast imaging described in this study is not optimal due to small number of elements and aperture size when compared to 2D imaging. This suboptimal setup results in poor signal to noise ratio in terms of Bmode image quality. However, past studies with a similar setup demonstrated the reliability of shear wave speed assessment both in isotropic phantoms [20] and in anisotropic media when compared to 2D [25].

Limitations also include the size and the cost of the 3D ultrafast ultrasound device which is in its current form difficult to translate to clinic. Technological strategies to reduce the number of channels and thus the size and the cost of the device could be implemented. Sparse arrays [26], micro-beamformers [27] embedded in the probe or row-column matrices [28] are all viable strategies to reach the goal of a clinical 3D ultrafast ultrasound device.

## V. CONCLUSION

We present in this study, 4D ultrafast natural shear wave depictions of the human heart, using a new method to assess natural shear wave velocities induced by valve closures. We visualized the complex propagation pattern for both aortic and mitral valve closure waves. Energy mapping and time of flight maps were computed to localize wave sources by visualizing isosurfaces. From time of flight maps, shear wave velocities were assessed along various directions from the localized sources. The proof of concept was successfully demonstrated on three healthy human volunteers. The method could enable mapping of ventricular stiffness for the diagnosis of cardiomyopathies and has the potential of becoming an important tool in the clinic.

ACKNOWLEDGMENT

This study was supported by the European Research Council under the European Union's Seventh Framework Program (FP/2007-2013) / ERC Grant Agreement n° 311025 and the ANR-10-IDEX-0001-02 PSL* Research University. We acknowledge the ART (Technological Research Accelerator) biomedical ultrasound program of INSERM. The Titan X Pascal used for this research was donated by the NVIDIA Corporation.